\documentclass[pre,twocolumn,showpacs]{revtex4}

\usepackage{hyperref}
\usepackage{graphicx}

\begin{document}

\title{Effective networks for real-time distributed processing}

\author{Gonzalo Travieso}
\email{gonzalo@ifsc.usp.br}

\author{Luciano da Fontoura Costa}
\email{luciano@ifsc.usp.br}

\affiliation{Instituto de F\'{\i}sica de S\~{a}o Carlos, 
  Universidade de S\~{a}o Paulo,
  Av. do Trabalhador S\~{a}o-Carlense 400,
  13566-590, S\~{a}o Carlos, SP, Brazil}

\begin{abstract}

  The problem of real-time processing is one of the most challenging
  current issues in computer sciences.  Because of the large amount of
  data to be treated in a limited period of time, parallel and
  distributed systems are required, whose performance depends on a
  series of factors including the interconnectivity of the processing
  elements, the application model and the communication protocol.  Given
  their flexibility for representing and modeling natural and
  human-made systems (such as the Internet and WWW), complex networks
  have become a primary choice in many research areas.  The current
  work presents how the concepts and methods of complex networks can
  be used to develop realistic models and simulations of distributed
  real-time system while taking into account two representative
  interconnection models: uniformly random and scale free
  (Barab\'asi-Albert), including the presence of background traffic of
  messages.  The interesting obtained results include the
  identification of the uniformly random interconnectivity scheme as
  being largely more efficient than the scale-free counterpart.

\end{abstract}


\maketitle

\section{Introduction}
\label{sec:intro}

We live in a world governed by action.  From the ample motion of our
planet to the intricacies of Brownian agitation, the universe is
pervaded by an endless flow of changes to which our lives are no
exception.  While little can originate from stillness, movement
imposes a continuing challenge to our senses.  An immediate and
important implication of movement is \emph{causality}, one of the most
essential elements in animal survival and also the key element in
scientific investigation.  In order to cope with such demands, animals
evolved an intricate neuronal `hardware' capable of analyzing moving
images at a high resolution and rate appropriate to enable an
immediate response, i.e.\ enough so as to favor their survival and
reproduction.  Such a type of reaction by dynamical systems is
technically known as
\emph{real-time} (e.g., \cite{Liu00}).  Despite the several advances in
computing technology achieved along the last decades, we still lag
well behind biological system as far as real-time processing and
recognition is concerned.  One of the possible ways to learn how to
develop automated systems for effective, real-time processing is to
look at the organization of biological systems for inspiration.
Another possibility is to model and simulate such systems in order to
try to identify particularly effective architectures and algorithms.
One of the fundamental organizational principles of biological
processing of information regards the inherent concurrency and
parallelism characterizing those systems.  Because neurons are
relatively slow in processing and transmitting information
(e.g., \cite{Kandel:2000}), high speed can only be achieved by
carefully interconnecting neurons so as to form groups or modules
working in parallel.  Indeed, the brain is currently known to be
organized according to interconnected modules~\cite{Zeki:2000}
resembling a distributed computer system.  In addition to the inherent
features of the modules and involved neuronal cells, one particular
feature of such modular processing systems concerns the specific way
in which the several components are \emph{interconnected}.

The interconnections between processing elements in a distributed
system can be natural and effectively represented in terms of
\emph{complex networks} (e.g., \cite{Albert_Barab:2002, Newman:2003, 
Boccaletti_etal:2006}), where each processor is associated to a node
while the interconnections between these nodes are expressed as edges.
Through such a simple analogy, it is possible to bridge the gap
between research in real-time distributed systems and the exciting
concepts, tools and results from the area of complex networks.
Although the origin of the latter area can be traced back to random
graphs (e.g., \cite{Bollobas:2001}), and despite their immediate
relationship with graph theory, the term \emph{complex network} has
been used to express the emphasis placed on graphs which exhibit
complex structured connectivity.\footnote{Although authors tend not to
consider uniformly random graphs (constant probability of connection
between any pair of nodes) as complex networks, even such simpler
structures do exhibit complex connections as a consequence of
statistical fluctuations.}

Thanks to technological advances in neuroanatomy and physiology, a
more comprehensive vision of neuronal interconnections underlying the
nervous systems of several animals is progressively emerging.  At the
microscopic --- cellular --- level, recent investigations have
suggested that neurons are interconnected through small world and even
scale free networks~\cite{Sporns:2000}. The macroscopic organization
of cortical areas~\cite{Zeki:2000} also seems to be organized
according to this principle~\cite{Sporns:2004}.

This article presents the application of complex networks as the means
for investigating the effect of alternative connectivity schemes,
namely uniformly random (i.e.\ Erd\H{o}s and R\'{e}nyi --- ER) and
scale free (i.e.\ Barab\'asi-Albert --- BA) network models, on the
overall performance of a real-time distributed processing system.
While the ER model represents the natural reference system for
connectivity, being almost universally considered as the null
hypothesis in complex network studies, the BA model is particularly
representative of natural --- including neuronal information
processing systems~\cite{Sporns:2000} --- and human-made systems such
as the Internet and the WWW~\cite{Albert_Barab:2002}.  ER networks
exhibit a characteristic node degree (i.e.\ the number of connections
of each node of the network), in the sense that their overall
connectivity can be well characterized in terms of the mean node
degree. Contrariwise, BA networks exhibit a power law distribution of
node degrees, which favors a heterogeneous connectivity, as well as
the appearance of hubs (e.g., \cite{Albert_Barab:2002,Newman:2003}).
In addition to their particular importance in modeling natural and
human-made systems, BA networks provide an interesting model for the
Internet and, consequently, grid computing systems --- an important
current trend in distributed computing \cite{FosterGrid} which
provides a good deal of the motivation for the present work.  In
addition, the consideration of the BA model allows us to investigate
the effect of the presence of hubs in parallel and grid systems --- as
implied by the Internet connectivity \cite{Faloutsos} --- on the
overall performance, as a counterpart to the otherwise almost regular
connectivity ensured by the ER model making it similar to uniform
parallel systems such as those involving mesh or hypercube
interconnectivity. However, unlike those architectures, ER (and BA)
networks are small-world.

The application model simulated here while considering these two
interconnecting models involves a master node which distributes tasks,
namely a stream of frames to be processed, among processing elements
in other nodes acting as clients according to their availability.
This assumes that both the source and destination of the frames are at
the same site. The processing protocol considers a communication model
involving routers connecting the clients to the master, as typically
found in practice.  Therefore, the overall modeling and simulation
approaches adopted in this work include many realistic elements common
to a real distributed processing system.

A previous work \cite{Costa_grid:2005} studied the effect of
interconnection topology in the performance of a grid computing
application. The application considered in that work involved the
processing of a number of not interacting tasks, with no real-time
requirements. The lack of real-time constraints reduces the importance
of traffic fluctuations enabling the use of average communication
times for performance evaluation. Correspondingly,
\cite{Costa_grid:2005} does not include traffic effects. Under the
real-time constraints of the application studied in the present work,
varying delays induced by traffic play a major role, and
application-independent traffic is thus included in the
simulations. Another reason for inclusion of traffic is that the
network topology strongly influences packet transit times under
traffic, as reported in many works
(e.g., \cite{Holme03,Tadic04,Zhao05,Tadic06}).

The current article starts by presenting the adopted network,
application and communication models, and follows by presenting
and discussing the obtained results.

\section{Models}

The model used for the simulations comprises three components: a
network model, a model for the communication between the collaborating
computers, and an application model. These models are described below.

\subsection{Network Models}

When studying complex networks, the interest can be focused on their
topologies, i.e.\ their structural properties, or on some dynamical
processes taking place in the network. In this work, we use complex
networks to describe the interconnection topology of a collection of
computers participating in a collaborative real-time computation. As
such, our main focus is on the dynamical processes of data
communication and computations in the computers interconnected through
the network.  In this work, two widely used network models are
considered: the Erd\H{o}s-R\'{e}nyi (ER) random network model with
fixed number of edges \cite{Erdos59} and the Barab\'{a}si-Albert (BA)
scale-free model \cite{Barabasi97}.

ER networks are constructed by considering $N$ isolated nodes (i.e.\
the network starts with no edges) and then adding edges one by one
between uniformly chosen pairs of nodes (avoiding duplicate
connections of nodes and self connections); the addition of edges is
repeated a pre-specified number $L$ of times. $N$ and $L$ are the
parameters of the ER model and the average degree is given by:
\begin{equation}
  \langle k \rangle = \frac{2L}{N}.\label{eq:kxl}
\end{equation}

BA networks are constructed starting with $m_0$ nodes and adding new
nodes one by one. When a new node is inserted, $m\le m_0$ edges are
added from this node to one of the previously existing nodes. The
nodes to be linked are chosen following the \emph{preferential
attachment} rule \cite{Barabasi97}.  The process is repeated until de
desired number of nodes $N$ is reached. In the simulations presented
in this work, the initial network was fixed with $m_0 = 2m+1$ fully
connected nodes. The parameters of the model are thus $N$ and $m$.
Considering that for each new node $m$ new edges are added and that
the initial network already has $m$ edges for each node, the total
number of edges is $L = mN$, and therefore
\begin{equation}
\langle k \rangle = 2m.\label{eq:kxm}
\end{equation}

\subsection{Application Model}

This work analyzes the influence of the network topology on real-time
collaborative computation. The computation considered here is defined
as follows: A special node in the network, called the
\emph{master}, is responsible for reading a stream of input data and
writing a stream of output data. The data arrives at the master in
packets, here called \emph{frames} in an analogy to real-time video
processing, at regular intervals and the result of their processing
must be output at the same interval.

For each input frame, an output frame is produced after the
realization of a certain amount of computation (the computational load
required for each frame). In this work, the load is considered equal
for all frames. The computation is not done by the master. Instead, a
collection of \emph{clients} book their willingness to participate in
the computation; when a new input frame arrives, the master chooses
one available client and sends the frame to it for processing.  After
receiving and processing the input frame, the client sends the output
frame back to the master; when the output frame arrives at the master,
the client that processed the frame is again registered as ready to
receive a new frame.

After arrival (or generation) at the master, each frame must be sent
to a client, processed and sent back to the master. As communication
delays in the network are unpredictable, the order of arrival of the
resulting frames at the master is not guaranteed to correspond to the
order in which they were originally delivered. To avoid output of the
frames out-of-order and also enable waiting for the transmission and
processing of the frames, a frame buffer must be maintained by the
master, where arriving frames are stored in the correct order. The
production of the output must then be delayed for some time, i.e.\ the
output of frames must start some time after the arrival of the first
frame. When a frame must be output, if it has not yet arrived it must
be dropped with resulting quality loss. It is therefore important to
allow sufficient time for the frames to arrive, but additional time
given to frame processing then implies in increased latency in the
production of the output. The time between the arrival of the first
frame and the start of the output (which is also the time each frame
will have available to be processed and returned to the master) is
henceforth quantified in terms of the number of frame intervals.

\subsection{Communication Model}

After a network is generated according to a given model and set of
parameters, its nodes are considered the \emph{routers} of a computer
network. The computers participating in the collaborative work are
\emph{hosts} connected to one of the routers. The master is connected
to a router randomly selected with uniform probability.  Not all of
the network participates in the computation. The number of
participating clients is a parameter of the simulation. Each client is
associated with a router selected with uniform probability, but a
limitation is imposed that each host (master or client) is associated
with a different router.  Only routers from the largest connected
component of the network are selected.

As the network is assumed not to be exclusively dedicated to the
frames computation, external traffic is simulated on the network by
the generation of packets between random pairs of routers.

After insertion in the network, the packets are routed from node to
node. The routers follow a ``shortest path'' routing strategy: each
router sends a packet to a neighboring router that strictly decreases
the number of steps remaining to reach the destination; if more than
one neighbor satisfies this condition, one of them is chosen at
random. While a router is routing and sending a packet to a neighbor,
it cannot handle other packets. Packets arriving during this operation
are queued in arrival order to be processed later; the routers are
assumed to have unbounded queuing capacity.

The time for processing and communication at each step on the network
is considered independent of the packet and router, although the
delivery time for different packets might differ due to queuing. If
the traffic in the network is low, the queues are empty or short, and
the time taken for a packet to reach the destination is proportional
to the topological distance between source and destination.  As the
traffic increases, congestion ensues \cite{Holme03,Zhao05,Tadic06},
and the delivery time grows to many times that of the uncongested
network.

\subsection{Parameters}

Here the model parameters and their values for the simulation results
described below are presented.

Both network models are characterized by two parameters: the number of
nodes $N$ and the number of edges (for the ER model) or number of
edges added for each new node (for the BA model). Henceforth the
latter parameters are represented by the average node degree $\langle
k \rangle$, that can be computed from the model parameters by using
equations~(\ref{eq:kxl}) and~(\ref{eq:kxm}).

The computation dynamics is described by the computational load for
the processing of each frame, the interval between frames and the
number of frames to wait before starting the output. Considering that
all clients are taken as identical (no difference in processing
power), the computational load can be given as the computational time
$T$ of the processing task. The time interval between frames will be
represented by $\tau$ and the number of frames to buffer by $B.$ The
output latency is therefore $B\tau.$

The time taken for a packet to traverse a step in the network from a
node to one of its neighbors, $h$, is the same for each packet and
router. As only the relation between the times are of importance, the
time scale is chosen such that $h=1$, the values of $T$ and $\tau$
being expressed in these units. The random traffic generation in the
network is assumed to be a Poisson process with inter-arrival times
given by an exponential distribution with average $1/(N\lambda);$ the
factor $N$ is introduced to make the amount of traffic proportional to
the size of the network; $\lambda$ is the per-node packet generation
frequency (in units compatible with $h=1$).

The remaining parameter is simply the number of clients $C \leq N-1.$
The parameters are listed in Table~\ref{tab:par}, together with their
range of values used in the simulations discussed below.

\begin{table}
  \caption{Model parameters and their values. Time and frequency
    parameter ``normalized'' units (see text); output start interval
    in number of frame intervals.}\label{tab:par}
  \begin{center}
    \begin{tabular}{ccc}
      \hline\\
      Parameter & Meaning & Values \\
      \hline\\
      $N$ & Number of nodes & $1000$ \\
      $\langle k \rangle$ & Average node degree & $2,6,10$\\
      $T$ & Frame computation time & $100$ \\
      $\tau$ & Frame interval & $5$\\
      $B$ & Output start interval & $10$--$50$\\
      $\lambda$ & Packet generation frequency & $0.001$--$0.02$ \\
      $C$ & Number of clients & $100$\\
      \hline
    \end{tabular}
  \end{center}
\end{table}

\section{Results and Discussion}

A computation is successful if all output frames are returned from the
clients and arrive at the master before they need for output. If a
frame arrives too late for output, that frame is dropped, and the
quality of the output is consequently reduced. Frames that arrive in
time are here called \emph{completed}.  The number of completed frames
is chosen as quality measure of the computation.  In the simulations,
a total of $1000$ frames needs to be computed.

\begin{figure*}
  \includegraphics[width=0.45\textwidth]{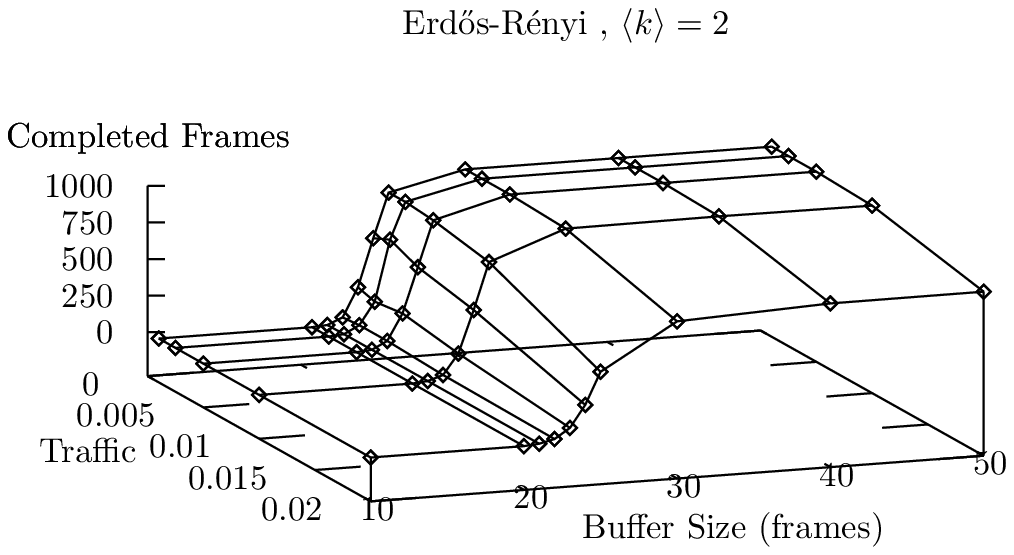}
  \includegraphics[width=0.45\textwidth]{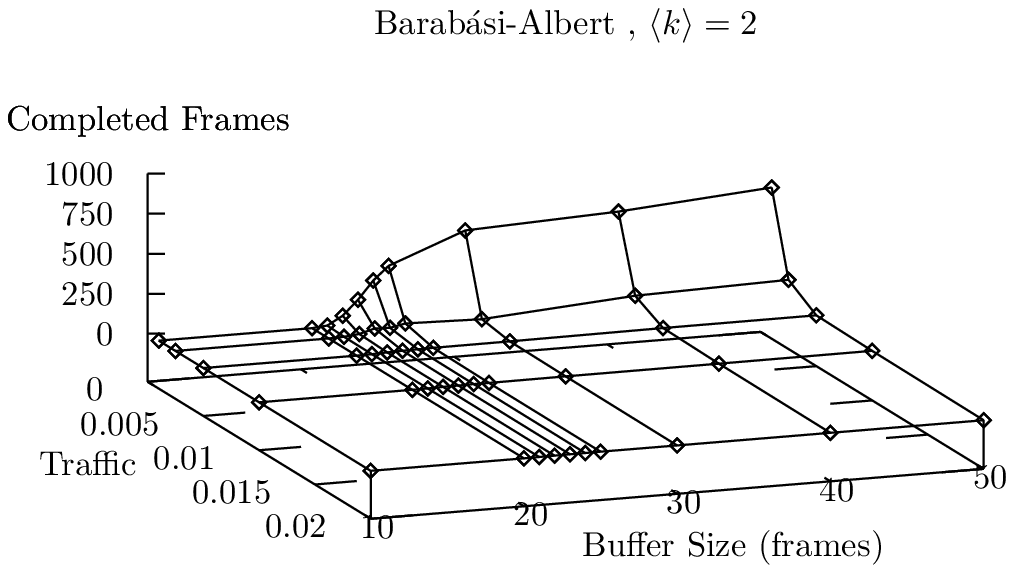}\\
  \hspace*{0.25\textwidth}(a)\hspace*{0.45\textwidth}(b)\vspace*{1cm}\\
  \includegraphics[width=0.45\textwidth]{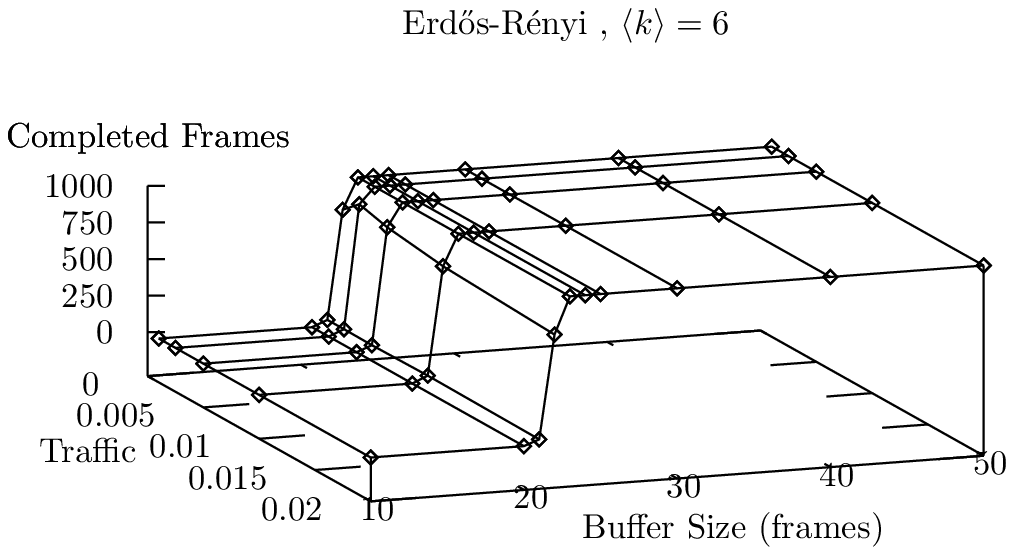}
  \includegraphics[width=0.45\textwidth]{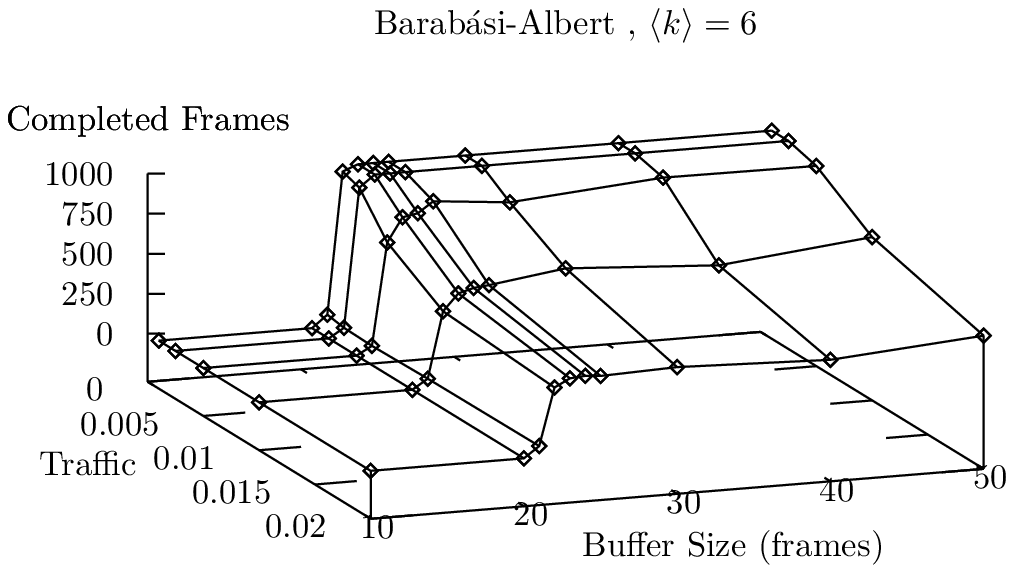}\\
  \hspace*{0.25\textwidth}(c)\hspace*{0.45\textwidth}(d)\vspace*{1cm}\\
  \includegraphics[width=0.45\textwidth]{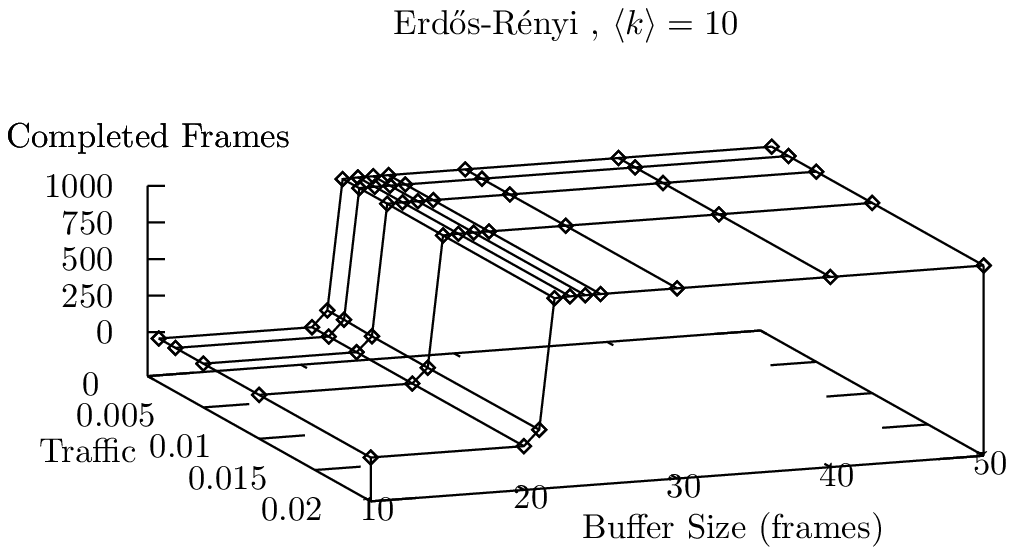}
  \includegraphics[width=0.45\textwidth]{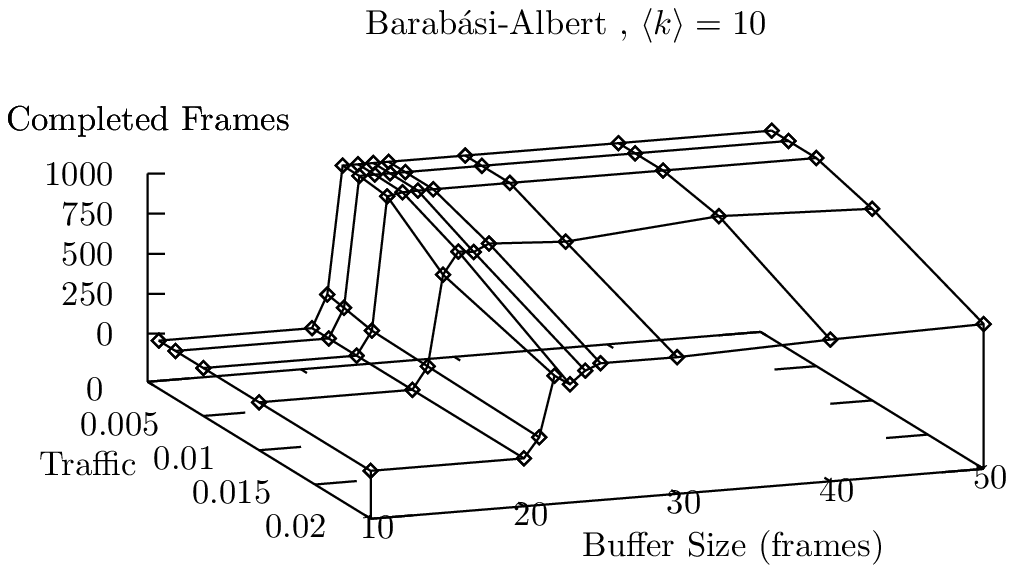}\\
  \hspace*{0.25\textwidth}(e)\hspace*{0.45\textwidth}(f)\\
  \caption{Number of completed frames for a total of 1000~frames as a
    function of network traffic and number of buffered frames, for
    ER networks (left column) and BA networks (right column). Model
    parameters are $N=1000,$ $T=100,$ $\tau=5,$ and $C=100.$}
  \label{fig:completed}
\end{figure*}

Figure~\ref{fig:completed} shows the number of completed frames as a
function of network traffic and the output latency, for ER and BA
networks of $1000$ nodes, with $\langle k \rangle = 2,6,10.$ The other
simulation parameters are: 100 clients, frame interval of $5$, frame
processing time of $100.$ The results shown are averages of $100$
simulations, each with a different network generated according to the
corresponding model and different traffic patterns.

Consider first the case of the Erd\H{o}s-R\'{e}nyi network with
$\langle k \rangle = 10$ (Fig.~\ref{fig:completed}(e)). This plot
shows a sharp transition on the number of completed frames for a
latency of about $20$ frame intervals. This transition is expected:
with $T=100$ and $\tau=5$, at least $T/\tau = 20$ frame intervals must
elapse before results start to arrive at the master. The fact that the
transition is sharp, close to this lower limit, and independent of
traffic in the studied region shows that an Erd\H{o}s-R\'{e}nyi
topology with $\langle k \rangle=10$ is efficient for this
application, that is, it introduces small delays. For $\langle k
\rangle = 6$ (Fig.~\ref{fig:completed}(c)), the results are similar,
but the transition is not so sharp and a larger latency is needed to
reach the plateau of all frames completed. In the case of $\langle k
\rangle = 2$ (Fig.~\ref{fig:completed}(a)), another effect appears: a
reduction on the number of completed frames occurs when the traffic is
increased.  The larger value of $B$ needed and the drop in the number
of completed frames with increased traffic for reduced values of
$\langle k \rangle$ are due to the reduction in the connectivity of
the network: Few edges connecting the nodes result in increased
average distances from the master to the clients; this affects the
time taken to deliver the frames and complete their calculations,
resulting in the need for an increase in the frame buffer and
therefore larger latency. Also, the presence of fewer edges means that
fewer alternative paths are available between the nodes, rising the
sensitivity of the network to increased traffic.

For the Barab\'{a}si-Albert networks, Figs.~\ref{fig:completed}(b),
(d), (f), the results show a much stronger influence of traffic. 
For $\langle k \rangle = 6$ and $\langle k \rangle = 10$ a continuous
drop of the number of completed frames is noticed as the amount of
traffic grows. For high traffic values, even large buffers are not
able to guarantee the completion of a sufficient amount of frames. For
$\langle k \rangle = 2$ the number of completed frames is small even
for reduced amounts of traffic.

\begin{figure*}
  \includegraphics[width=0.45\textwidth]{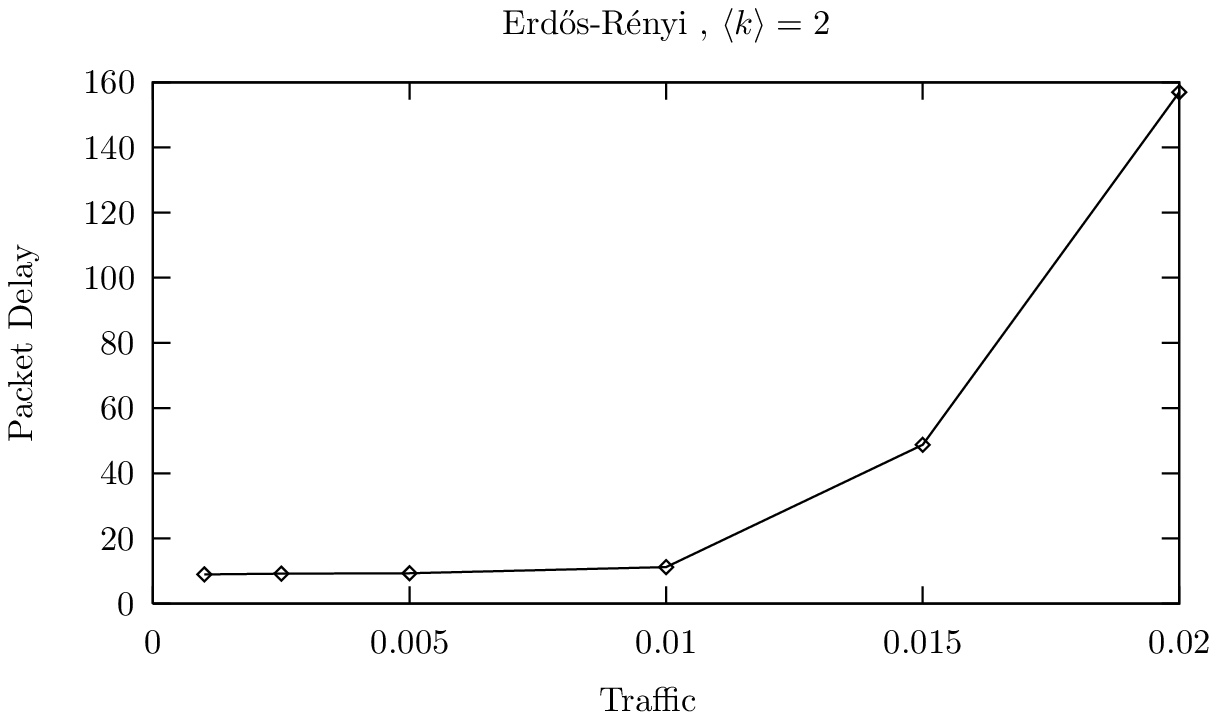}
  \includegraphics[width=0.45\textwidth]{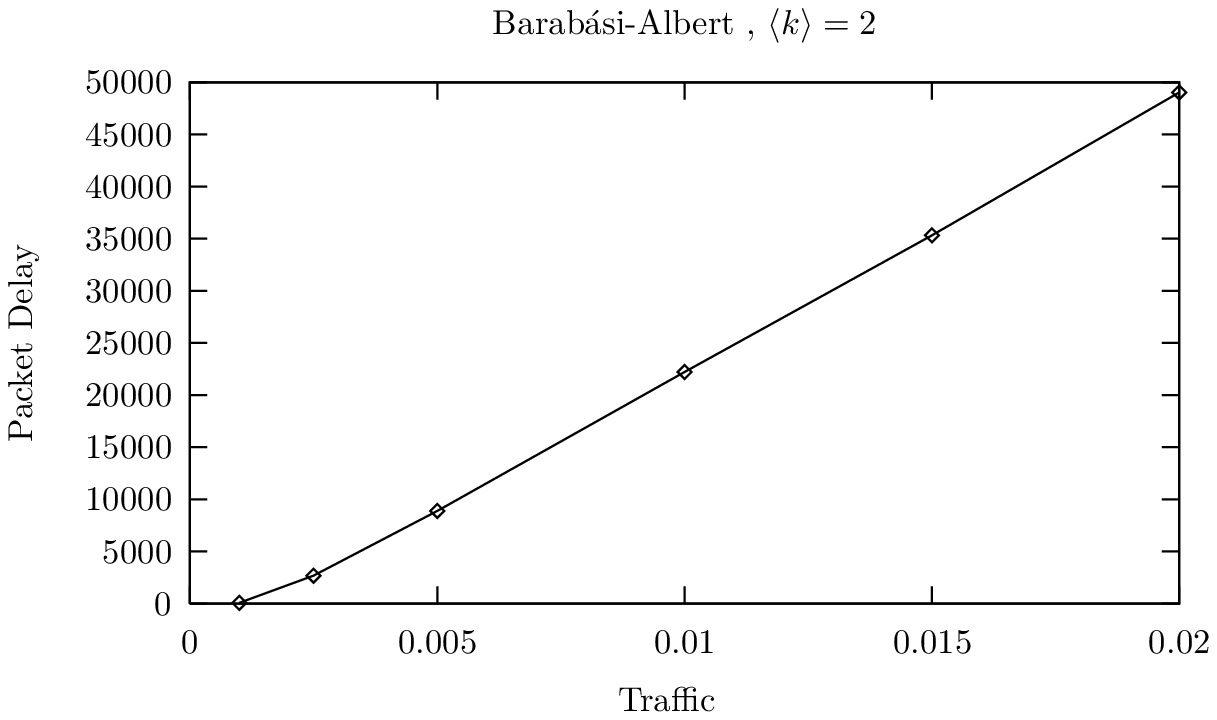}\\
  \hspace*{0.25\textwidth}(a)\hspace*{0.45\textwidth}(b)\vspace*{1cm}\\
  \includegraphics[width=0.45\textwidth]{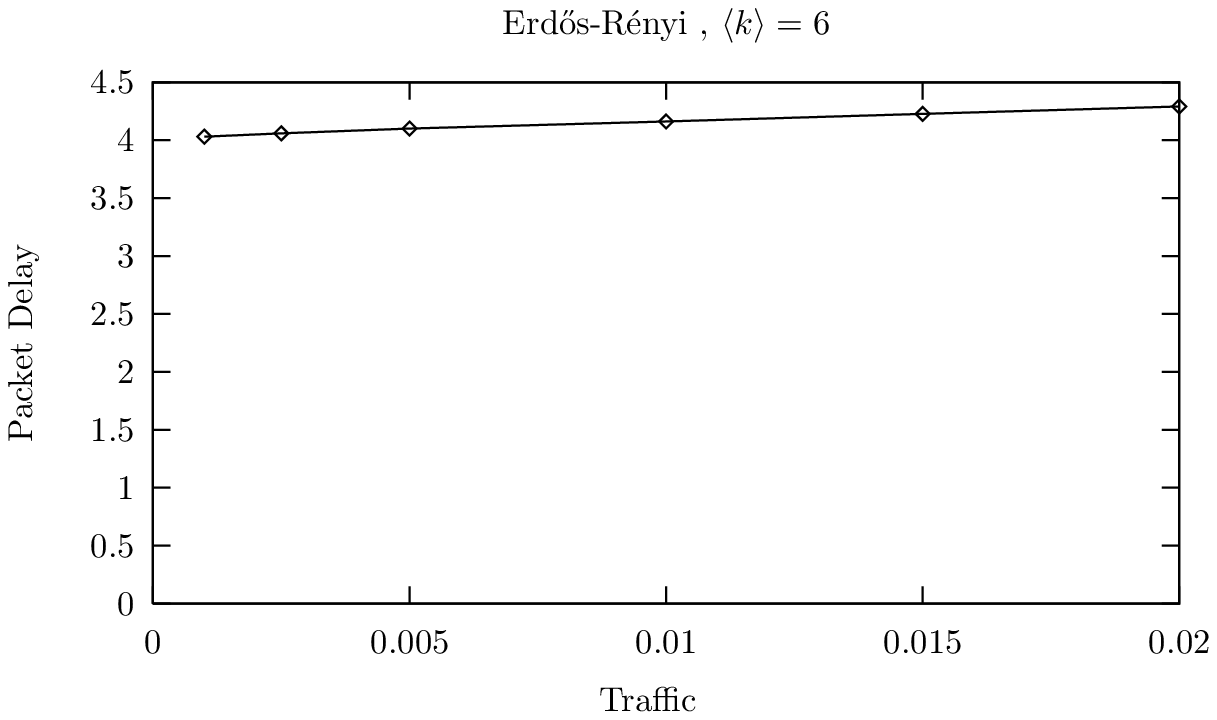}
  \includegraphics[width=0.45\textwidth]{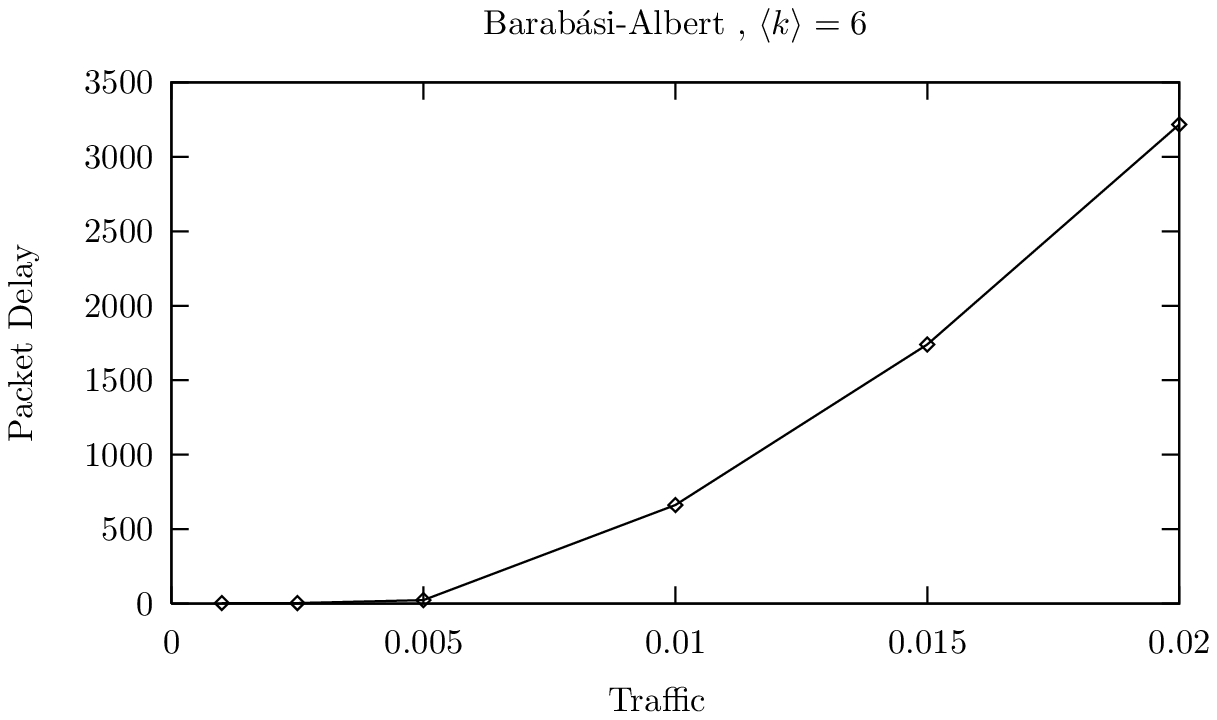}\\
  \hspace*{0.25\textwidth}(c)\hspace*{0.45\textwidth}(d)\vspace*{1cm}\\
  \includegraphics[width=0.45\textwidth]{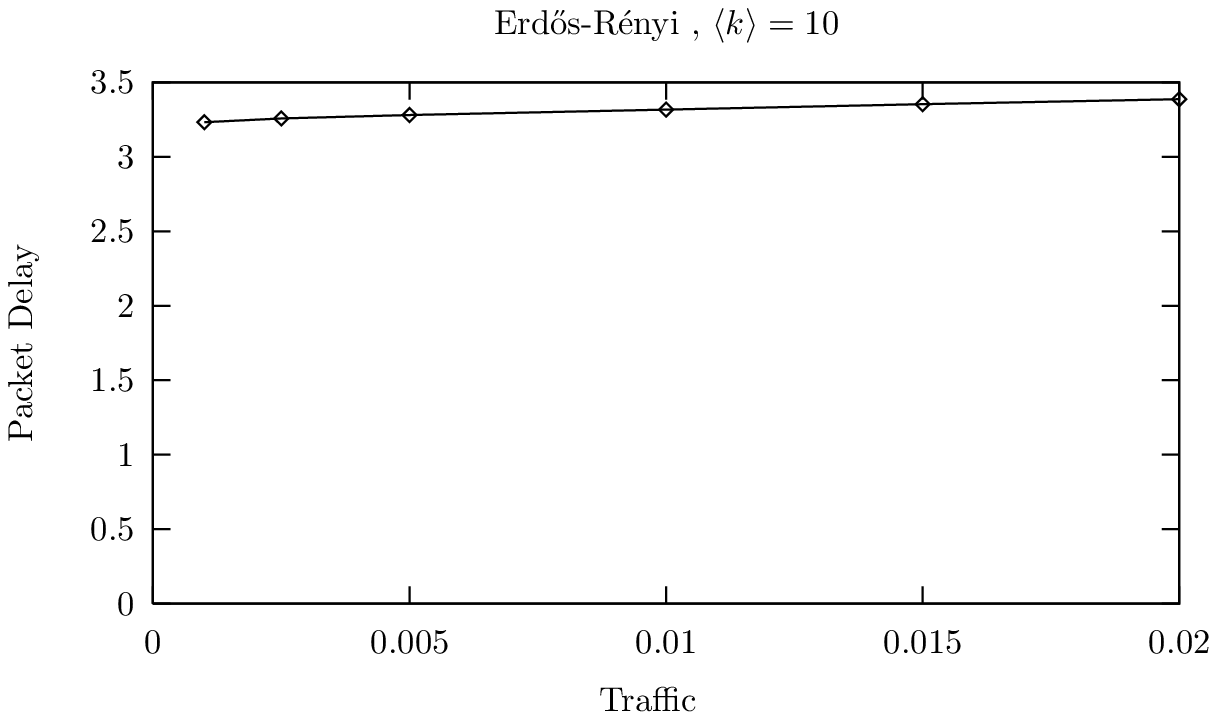}
  \includegraphics[width=0.45\textwidth]{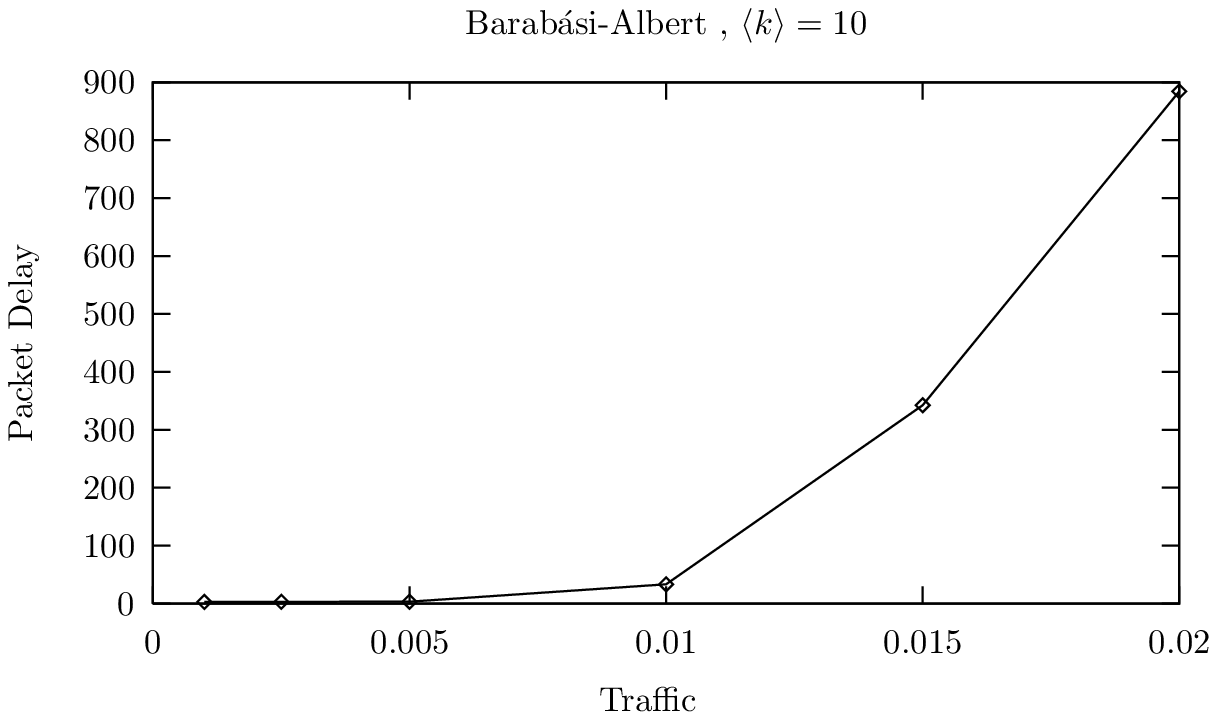}\\
  \hspace*{0.25\textwidth}(e)\hspace*{0.45\textwidth}(f)\\
  \caption{Average delay for the delivery of packets in the network
    (time taken by the packets from source to destination) as a
    function of network traffic. All packets in the network are
    included in the average (not only packets that transport frames).
    Model parameters are $N=1000,$ $T=100,$ $B=30,$ $\tau=5,$ and
    $C=100.$}
  \label{fig:traffic}
\end{figure*}

In order to better understand these results, Figure~\ref{fig:traffic}
plots the average packet transmission delay for the same situations as
presented in Figure~\ref{fig:completed}. The delay is computed as the
time taken from the delivery of a packet at the source to the arrival
at the destination. As shortest path routing is used and the time
taken at each step (hop) is unitary, the average delay should equal
the average distance between nodes under reduced traffic. This can be
seen for the ER networks with $\langle k \rangle = 6$ and $\langle k
\rangle = 10$ (Figs.~\ref{fig:traffic}(c),(e)), where the graphs are
flat with a delay value about the value of the average distances. A
different behavior is seen for ER networks with $\langle k \rangle =
2$ (Fig.~\ref{fig:traffic}(a)). At a packet generation frequency of
about $\lambda = 0.01$, the delay starts to grow linearly with the
amount of traffic. This is due to the onset of congestion in the
network: some nodes start receiving packets far more frequently than
they can handle, leading to increased queuing times of the packets in
the nodes. After congestion, the average delays grow fast to many
orders of magnitude of the average distance.
Figure~\ref{fig:traffic}(f) shows that congestion occurs for the BA
network with $\langle k \rangle = 10$ for a similar value of $\lambda
= 0.01$, but note that the increase in delay is steeper after that
point. For $\langle k \rangle = 6,$ BA networks display congestion at
lower traffics (about $\lambda=0.005$) and even steeper increases of
delay. The problem is accentuated for $\langle k \rangle = 2$
(Fig.~\ref{fig:traffic}(b)), where congestion occurs even for small
amounts of traffic.

The reason for this greater sensitivity of the Barab\'{a}si-Albert
networks to traffic in comparison with the Erd\H{o}s-R\'{e}nyi
counterparts can be easily understood. In fact, the preferential
attachment rule of BA networks induces the creation of nodes with a
high degree (hubs). Due to their high connectivity, these hubs appear
in many of the shortest paths of the network. Although hubs are
created, the number of hubs is always small, and most of the nodes
have small connectivity and take part in just a few shortest
paths. Therefore, a few nodes of the network become responsible for
routing almost all of the traffic, resulting in large packet queues
and congestion in these nodes. The lower the total connectivity of the
network, the more pronounced is this problem, as fewer links imply
fewer alternative shortest paths. ER networks, on the other hand,
distribute the connectivity homogeneously between all nodes, thus
generating a better distribution of shortest paths among the nodes of
the network.

\begin{figure*}
  \includegraphics[width=0.45\textwidth]{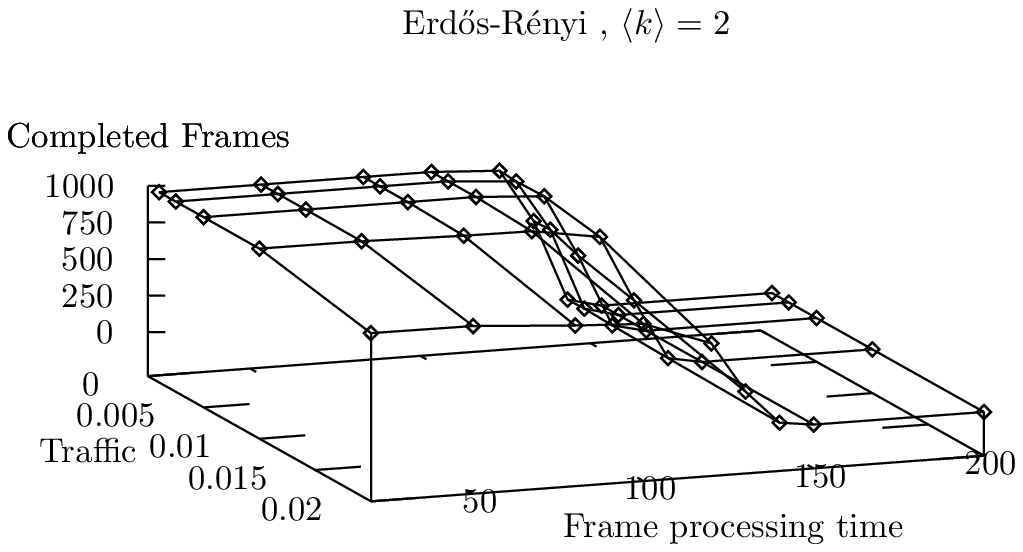}
  \includegraphics[width=0.45\textwidth]{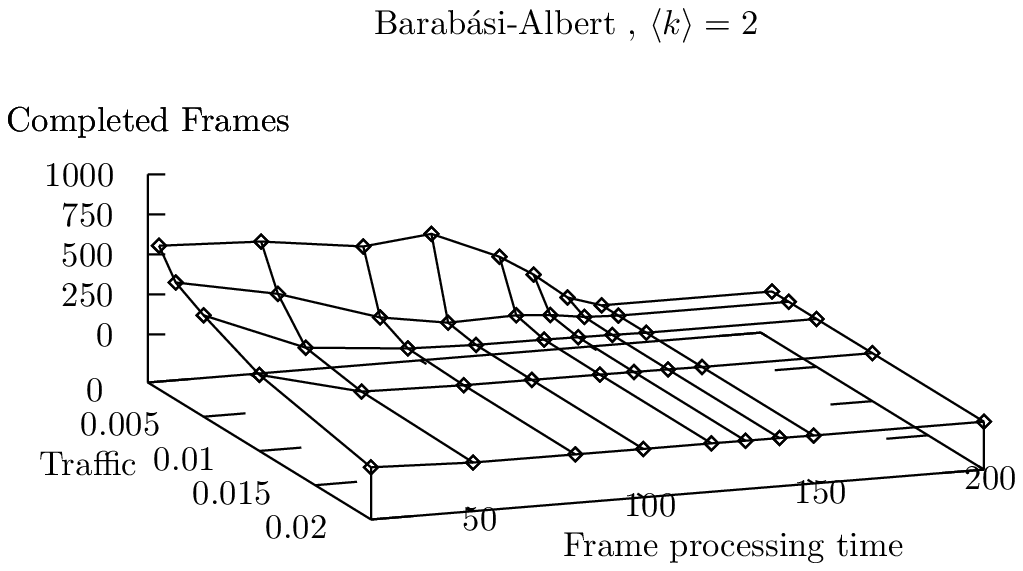}\\
  \hspace*{0.25\textwidth}(a)\hspace*{0.45\textwidth}(b)\vspace*{1cm}\\
  \includegraphics[width=0.45\textwidth]{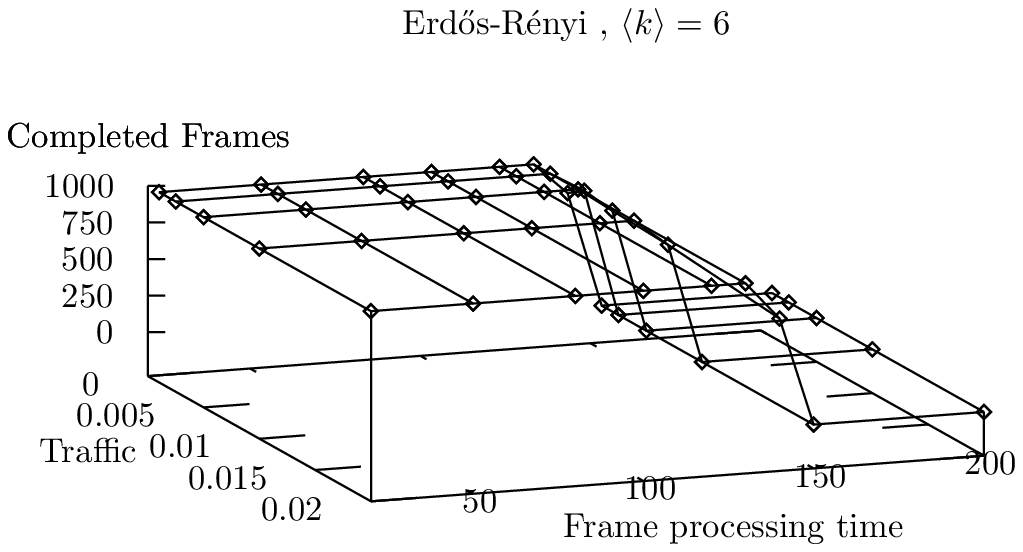}
  \includegraphics[width=0.45\textwidth]{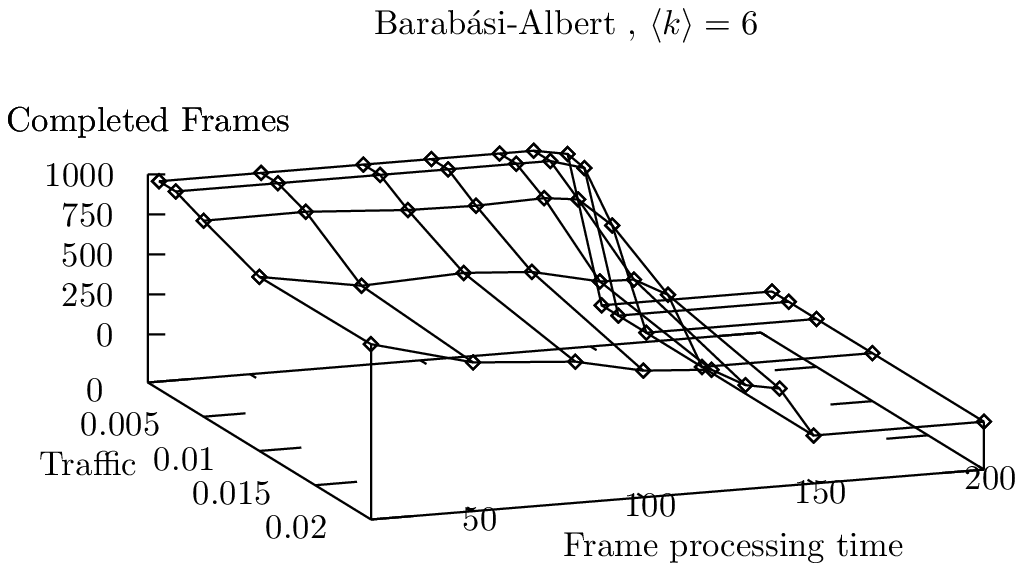}\\
  \hspace*{0.25\textwidth}(c)\hspace*{0.45\textwidth}(d)\vspace*{1cm}\\
  \includegraphics[width=0.45\textwidth]{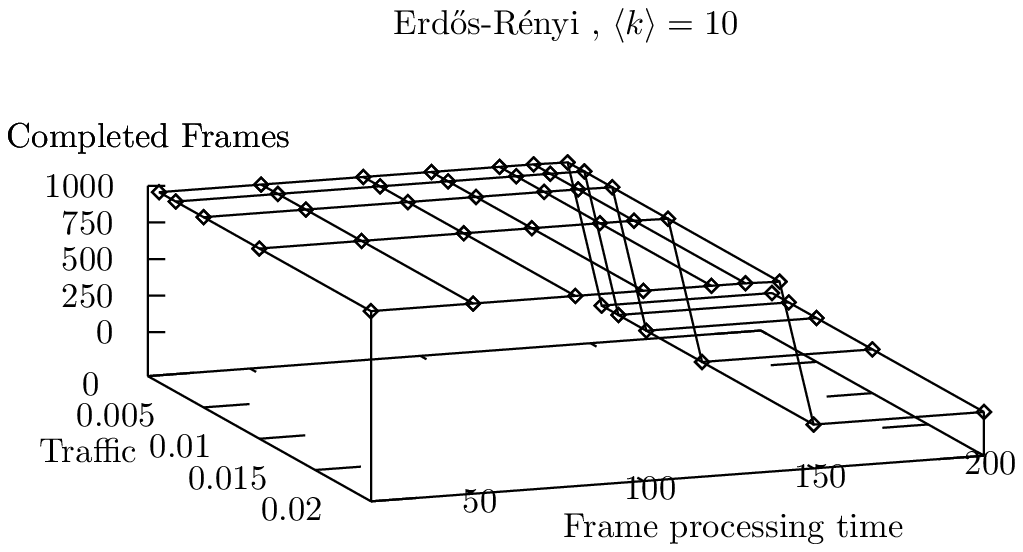}
  \includegraphics[width=0.45\textwidth]{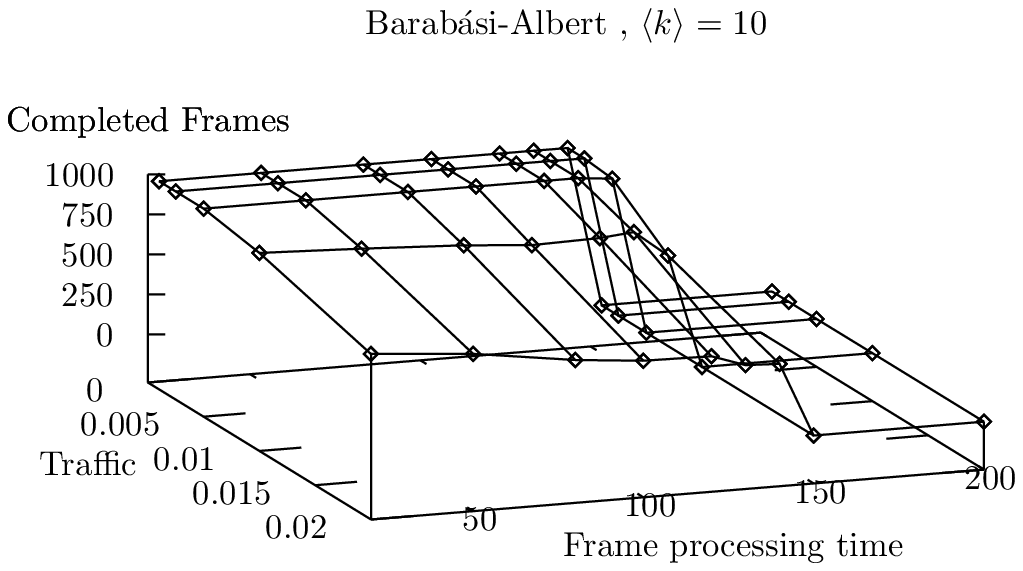}\\
  \hspace*{0.25\textwidth}(e)\hspace*{0.45\textwidth}(f)\\
  \caption{Number of completed frames from a total of 1000~frames as a
    function of network traffic and frame computation time, for
    ER networks (left column) and BA networks (right column). Model
    parameters are $N=1000,$ $B=30,$ $\tau=5,$ and $C=100.$}
  \label{fig:load}
\end{figure*}

To assess the influence of the computational load associated with each
frame (parameter $T$), Figure~\ref{fig:load} shows the number of
completed frames as a function of traffic and frame processing time.
For ER networks with $\langle k \rangle=6$ and $\langle k \rangle=10$,
where no congestion occurs, two plateaux, one with all frames
completed and the other with no frames completed, with a sharp
transition between $T=140$ and $T=150,$ are clearly seen. For the
other cases, where traffic is important, a gradual decay of the number
of completed frames is seen for increased traffic, as already seen in
Figure~\ref{fig:completed}, but there is also a gradual decrease of
the number of completed frames as the frame computation time increases
(before the transition to the no completion plateau). The higher the
traffic, the steeper is the decrease of the number of completed frames
with frame completion time.

The above results can be understood by the following reasoning. After
the generation of a frame $f$, it must be delivered to a client,
processed, and sent back to the master. Total processing time for $f$,
$P(f)$ is given by
\begin{equation}
	P(f) = w(f) + t_{mc}(f) + t_{cm}(f') + T
	\label{eq:proctime}
\end{equation}
where $w(f)$ if the time $f$ waits for a ready client, $t_{mc}(f)$ is
the travel time of $f$ from master to client, and $t_{cm}(f')$ is the
travel time from client to master of the frame generated by the
processing of $f$. Travel times $t_{mc}(f)$ and $t_{cm}(f')$ are
generally different (although the topological distance is the same in
both directions) due to possibly different traffic conditions at the
two transit periods. The condition for the completion in time of $f$
is that $P(f)$ is less then the accepted latency $B\tau$, giving
\begin{equation}
	w(f) + t_{mc}(f) + t_{cm}(f') + T \le B\tau.
	\label{eq:framecond}
\end{equation}
This condition must be satisfied by most frames.  Under low traffic
conditions, $t_{mc}(f)\approx t_{cm}(f')$ is close to the topological
distance between master and client and small due to the small world
property of the network models used, and the buffer used can be small,
implying small latencies.  Under heavy traffic, transit times can be
very large (see Fig.~\ref{fig:traffic}), resulting in the need of high
values of $B$ and therefore large latencies; also, even with large
$B$, the fluctuations in traffic are high, and many frames will be
lost.  This renders the distributed system useless for the
application.

\begin{figure*}
  \centering
  \includegraphics[width=0.45\textwidth]{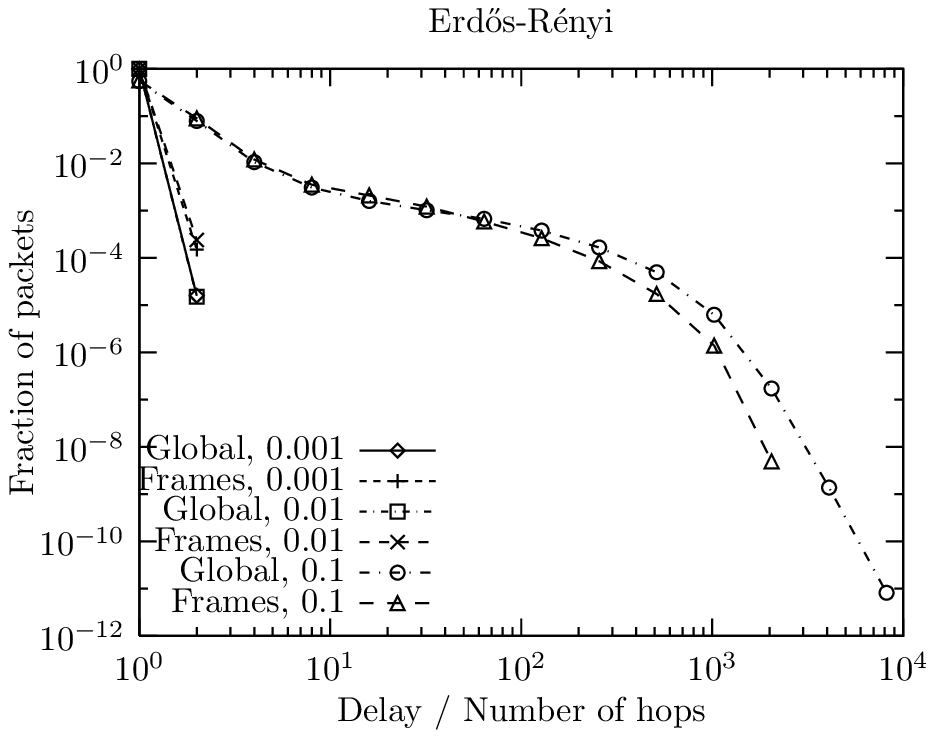}(a)
  \includegraphics[width=0.45\textwidth]{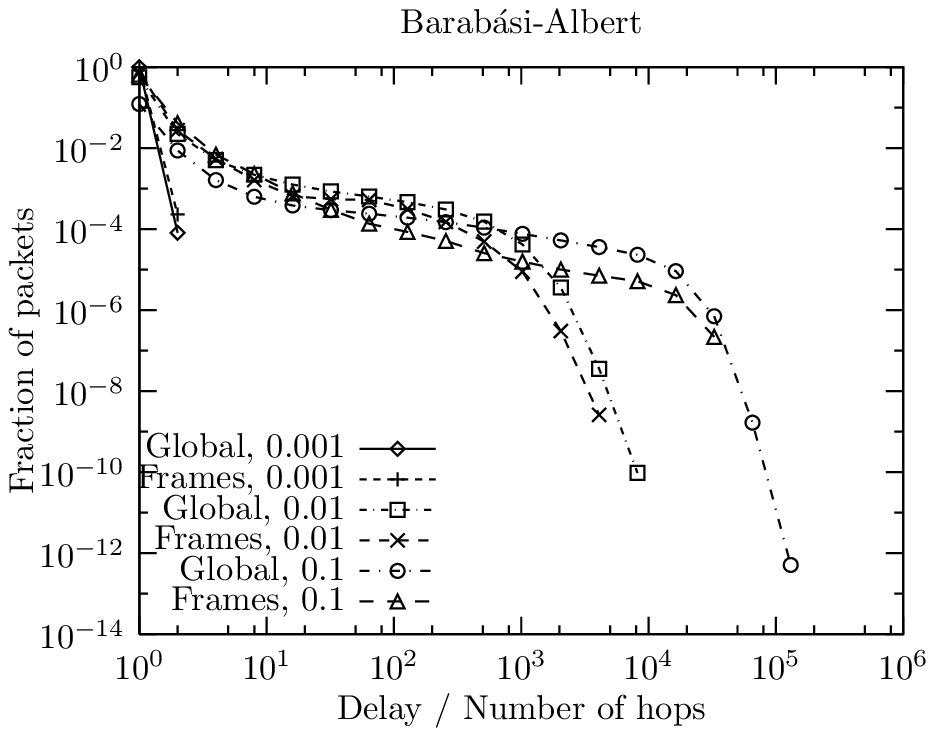}(b)
  \includegraphics[width=0.45\textwidth]{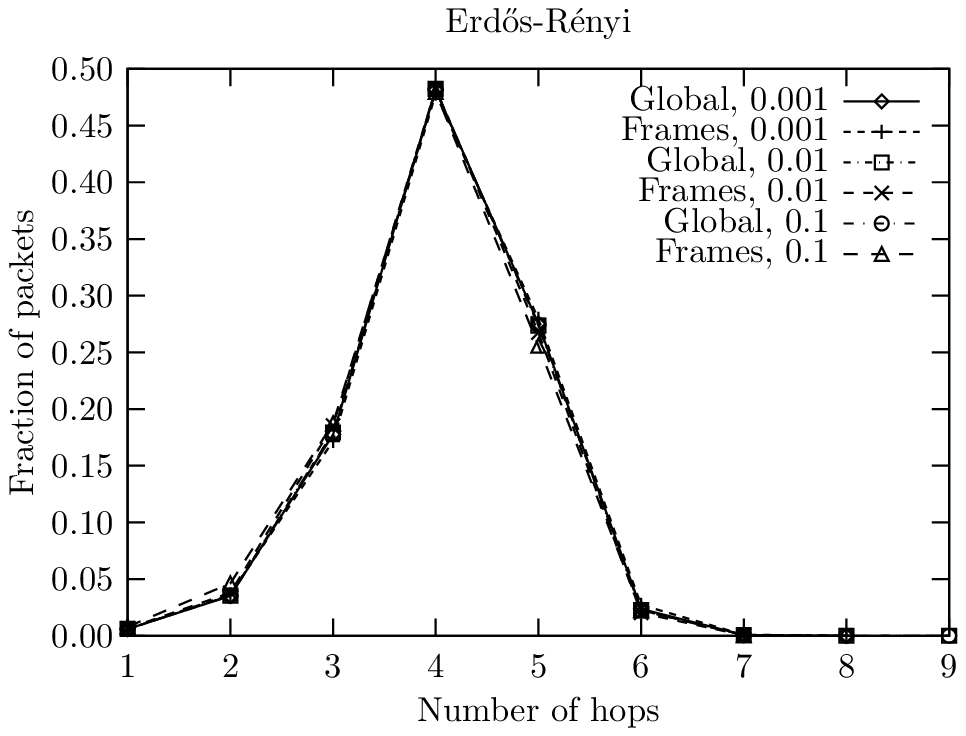}(c)
  \includegraphics[width=0.45\textwidth]{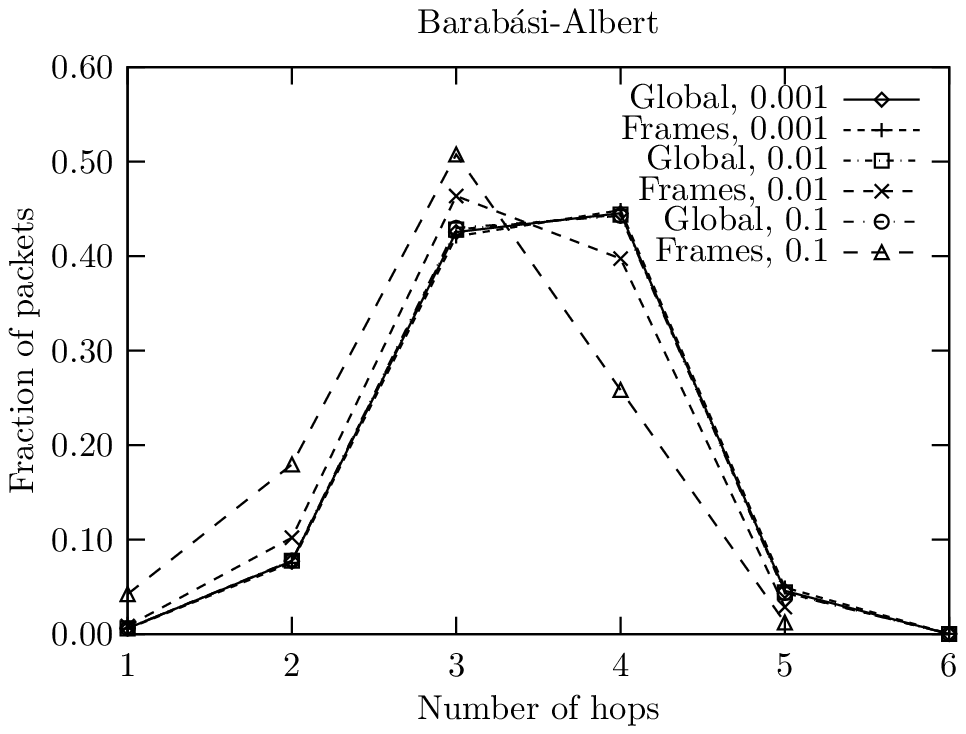}(d)
  \caption{Communication properties of frames (Frames) as compared to all
    packets (Global) in the network: per hop delays (a) and (b) and
    number of hops (c) and (d), for the Erd\H{o}s-R\'{e}nyi (a) and
    (c) and Barab\'{a}si-Albert (b) and (d) network models. Simulation
    parameters are as in Table~\ref{tab:par} and $\langle k
    \rangle=6,$ $B=50,$ and $\lambda=0.001, 0.01, 0.1.$}
  \label{fig:fxp}
\end{figure*}

When a client returns the result of a computation to the master, it is
automatically registered as able to receive a new packet.  Therefore,
it is reasonable to suppose that clients that communicate faster with
the master will receive a larger number of frames to compute and as a
result the delays associated with the communication of frames can be
smaller than the network averages.  Also, as nearby clients tend to
communicate faster with the master, the average distances (number of
hops) traveled by frames can be lower than for the other packets.
Figures.~\ref{fig:fxp}(a) and~(b) show the distribution of packets
with given per-hop delays (i.e., the ratio of the packet delay to the
number of hops traversed by the packet) for the two considered network
models and three different traffic conditions.  It can be seen that
under heavy traffic, specially for the Barab\'{a}si-Albert model
(which is more sensitive to traffic), the frames are subjected to
smaller delays than the other packets, but differences are substantial
only for a small number of packets with prohibitively large delays,
resulting in no advantage for the computation.  The distribution of
the number of hops traveled by all packet and the frames, for the two
network models, are shown in Figures~\ref{fig:fxp}(c) and~(d).  They
show that the clients effectively participating in the computation are
uniformly distributed among the network nodes, with a perceptible
change in distribution only for high traffic condition in the
Barab\'{a}si-Albert network model; in this latter case, effectively
operating clients are positioned closer to the master node, implying
that farther client nodes are receiving a smaller number of frames to
be computed; again no advantage comes to the application being
processed, as the difference occurs only in a traffic condition where
the number of droped frames is too high.

\section{Concluding Remarks}

Combined with the availability of ever increasing amounts of data, the
continuing advances scientific simulations have imposed serious
demands for real-time processing.  A natural means to cope with such a
pressure is to develop and apply distributed systems, including the
possibility of learning from biological systems and the application of
Internet-based grid computing.  Because of the high cost in
implementing such solutions, it becomes essential to have access to
realistic and effective modeling and simulation methodologies.  The
current work has described how concepts and methods from the modern
area of complex networks research can be effectively applied in order
to model and simulate with a good level of realism distributed systems
for real-time processing, with emphasis focused on grid computing
structures with connectivity underlined by the Internet.  At the same
time, because the BA model reflects some important connectivity
features found in neuronal processing systems, the development and
evaluation of such complex network models for real-time processing
bear potential implications also for understanding biological
processing.

Given its compatibility with some Internet topological features, and
also because of its potential compatibility with neuronal processing
systems, the Barab\'{a}si-Albert complex network model has been
selected in order to define the overall connectivity of the
distributed real-time processing system.  The Erd\H{o}s-R\'{e}nyi
complex network model was also considered as a null hypothesis
characterized by a high uniformity of node degree.  Realistic models
were assumed for the application and communication dynamics, including
the effect of background message traffic, while the overall
performance was quantified in terms of the total number of processed
frames with respect to varying traffic intensity, buffer size and
frame processing time.  The obtained results included the
identification of critical parameter configurations which are closely
related to the model parameters and overall connectivity.  Of special
interest is the clear superiority of ER networks over BA networks.
This is a result of the better handling of traffic by ER networks, as
a consequence of the better distribution of connectivity between the
nodes.

Possible future works include the consideration of other complex
network models and applications, as well as inclusion of variability
in the computing power of the clients, in the processing requirements
of different frames and in the communication times between nodes. It
is also of particular interest to study the effect of different
routing algorithms and packet queuing strategies at the routers.

\begin{acknowledgments}
Luciano da F. Costa thanks FAPESP (05/00587-5) and CNPq (308231/03-1)
for sponsorship.  The work of G. Travieso was partially supported by
FAPESP (03/08269-7).
\end{acknowledgments}

\bibliography{rt}

\begin{thebibliography}{18}
\expandafter\ifx\csname natexlab\endcsname\relax\def\natexlab#1{#1}\fi
\expandafter\ifx\csname bibnamefont\endcsname\relax
  \def\bibnamefont#1{#1}\fi
\expandafter\ifx\csname bibfnamefont\endcsname\relax
  \def\bibfnamefont#1{#1}\fi
\expandafter\ifx\csname citenamefont\endcsname\relax
  \def\citenamefont#1{#1}\fi
\expandafter\ifx\csname url\endcsname\relax
  \def\url#1{\texttt{#1}}\fi
\expandafter\ifx\csname urlprefix\endcsname\relax\def\urlprefix{URL }\fi
\providecommand{\bibinfo}[2]{#2}
\providecommand{\eprint}[2][]{\url{#2}}

\bibitem[{\citenamefont{Liu}(2000)}]{Liu00}
\bibinfo{author}{\bibfnamefont{J.~W.~S.} \bibnamefont{Liu}},
  \emph{\bibinfo{title}{Real-Time Systems}}
  (\bibinfo{publisher}{Prentice-Hall}, \bibinfo{year}{2000}).

\bibitem[{\citenamefont{Kandel et~al.}(2000)\citenamefont{Kandel, Schwartz, and
  Jessell}}]{Kandel:2000}
\bibinfo{author}{\bibfnamefont{E.~R.} \bibnamefont{Kandel}},
  \bibinfo{author}{\bibfnamefont{J.~H.} \bibnamefont{Schwartz}},
  \bibnamefont{and} \bibinfo{author}{\bibfnamefont{T.~M.}
  \bibnamefont{Jessell}}, \emph{\bibinfo{title}{Principles of Neural Science}}
  (\bibinfo{publisher}{McGraw-Hill}, \bibinfo{year}{2000}).

\bibitem[{\citenamefont{Zeki}(2000)}]{Zeki:2000}
\bibinfo{author}{\bibfnamefont{S.}~\bibnamefont{Zeki}},
  \emph{\bibinfo{title}{Inner Vision: An Exploration of Art and the Brain}}
  (\bibinfo{publisher}{Oxford University Press}, \bibinfo{year}{2000}).

\bibitem[{\citenamefont{Albert and Barab\'asi}(2002)}]{Albert_Barab:2002}
\bibinfo{author}{\bibfnamefont{R.}~\bibnamefont{Albert}} \bibnamefont{and}
  \bibinfo{author}{\bibfnamefont{A.~L.} \bibnamefont{Barab\'asi}},
  \bibinfo{journal}{Rev. Mod. Phys.} \textbf{\bibinfo{volume}{74}},
  \bibinfo{pages}{47} (\bibinfo{year}{2002}).

\bibitem[{\citenamefont{Newman}(2003)}]{Newman:2003}
\bibinfo{author}{\bibfnamefont{M.~E.~J.} \bibnamefont{Newman}},
  \bibinfo{journal}{SIAM Review} \textbf{\bibinfo{volume}{45}},
  \bibinfo{pages}{167} (\bibinfo{year}{2003}),
  \bibinfo{note}{cond-mat/0303516}.

\bibitem[{\citenamefont{Boccaletti et~al.}(2006)\citenamefont{Boccaletti,
  Latora, Moreno, Chavez, and Hwang}}]{Boccaletti_etal:2006}
\bibinfo{author}{\bibfnamefont{S.}~\bibnamefont{Boccaletti}},
  \bibinfo{author}{\bibfnamefont{V.}~\bibnamefont{Latora}},
  \bibinfo{author}{\bibfnamefont{Y.}~\bibnamefont{Moreno}},
  \bibinfo{author}{\bibfnamefont{M.}~\bibnamefont{Chavez}}, \bibnamefont{and}
  \bibinfo{author}{\bibfnamefont{D.-U.} \bibnamefont{Hwang}},
  \bibinfo{journal}{Physics Reports}  (\bibinfo{year}{2006}),
  \bibinfo{note}{accepted}.

\bibitem[{\citenamefont{Bollob\'as}(2001)}]{Bollobas:2001}
\bibinfo{author}{\bibfnamefont{B.}~\bibnamefont{Bollob\'as}},
  \emph{\bibinfo{title}{Random Graphs}} (\bibinfo{publisher}{Cambridge
  University Press}, \bibinfo{year}{2001}).

\bibitem[{\citenamefont{Sporns et~al.}(2000)\citenamefont{Sporns, Tononi, and
  Edelman}}]{Sporns:2000}
\bibinfo{author}{\bibfnamefont{O.}~\bibnamefont{Sporns}},
  \bibinfo{author}{\bibfnamefont{G.}~\bibnamefont{Tononi}}, \bibnamefont{and}
  \bibinfo{author}{\bibfnamefont{G.~M.} \bibnamefont{Edelman}},
  \bibinfo{journal}{Cerebral Cortex} \textbf{\bibinfo{volume}{10}},
  \bibinfo{pages}{127} (\bibinfo{year}{2000}).

\bibitem[{\citenamefont{Sporns et~al.}(2004)\citenamefont{Sporns, Chialvo,
  Kaiser, and Hilgetag}}]{Sporns:2004}
\bibinfo{author}{\bibfnamefont{O.}~\bibnamefont{Sporns}},
  \bibinfo{author}{\bibfnamefont{D.}~\bibnamefont{Chialvo}},
  \bibinfo{author}{\bibfnamefont{M.}~\bibnamefont{Kaiser}}, \bibnamefont{and}
  \bibinfo{author}{\bibfnamefont{C.~C.} \bibnamefont{Hilgetag}},
  \bibinfo{journal}{Trends in Cognitive Sciences} \textbf{\bibinfo{volume}{8}},
  \bibinfo{pages}{418} (\bibinfo{year}{2004}).

\bibitem[{\citenamefont{Foster et~al.}(2001)\citenamefont{Foster, Kesselman,
  and Tuecke}}]{FosterGrid}
\bibinfo{author}{\bibfnamefont{I.}~\bibnamefont{Foster}},
  \bibinfo{author}{\bibfnamefont{C.}~\bibnamefont{Kesselman}},
  \bibnamefont{and} \bibinfo{author}{\bibfnamefont{S.}~\bibnamefont{Tuecke}},
  \bibinfo{journal}{International Journal of High Performance Computing
  Applications} \textbf{\bibinfo{volume}{15}}, \bibinfo{pages}{200}
  (\bibinfo{year}{2001}).

\bibitem[{\citenamefont{Faloutsos et~al.}(1999)\citenamefont{Faloutsos,
  Faloutsos, and Faloutsos}}]{Faloutsos}
\bibinfo{author}{\bibfnamefont{M.}~\bibnamefont{Faloutsos}},
  \bibinfo{author}{\bibfnamefont{P.}~\bibnamefont{Faloutsos}},
  \bibnamefont{and}
  \bibinfo{author}{\bibfnamefont{C.}~\bibnamefont{Faloutsos}}, in
  \emph{\bibinfo{booktitle}{SIGCOMM '99: Proceedings of the conference on
  applications, technologies, architectures, and protocols for computer
  communication}} (\bibinfo{year}{1999}), pp. \bibinfo{pages}{251--262}.

\bibitem[{\citenamefont{da~F.~Costa et~al.}(2005)\citenamefont{da~F.~Costa,
  Travieso, and Ruggiero}}]{Costa_grid:2005}
\bibinfo{author}{\bibfnamefont{L.}~\bibnamefont{da~F.~Costa}},
  \bibinfo{author}{\bibfnamefont{G.}~\bibnamefont{Travieso}}, \bibnamefont{and}
  \bibinfo{author}{\bibfnamefont{C.~A.} \bibnamefont{Ruggiero}},
  \bibinfo{journal}{European Physical Journal B} \textbf{\bibinfo{volume}{44}},
  \bibinfo{pages}{119} (\bibinfo{year}{2005}),
  \bibinfo{note}{cond-mat/0312603}.

\bibitem[{\citenamefont{Holme}(2003)}]{Holme03}
\bibinfo{author}{\bibfnamefont{P.}~\bibnamefont{Holme}},
  \bibinfo{journal}{Advances in Complex Systems} \textbf{\bibinfo{volume}{6}},
  \bibinfo{pages}{163} (\bibinfo{year}{2003}).

\bibitem[{\citenamefont{Tadi\'{c} et~al.}(2004)\citenamefont{Tadi\'{c},
  Thurner, and Rodgers}}]{Tadic04}
\bibinfo{author}{\bibfnamefont{B.}~\bibnamefont{Tadi\'{c}}},
  \bibinfo{author}{\bibfnamefont{S.}~\bibnamefont{Thurner}}, \bibnamefont{and}
  \bibinfo{author}{\bibfnamefont{G.~J.} \bibnamefont{Rodgers}},
  \bibinfo{journal}{Physical Review E} \textbf{\bibinfo{volume}{69}},
  \bibinfo{pages}{036102} (\bibinfo{year}{2004}).

\bibitem[{\citenamefont{Zhao et~al.}(2005)\citenamefont{Zhao, Lai, Park, and
  Ye}}]{Zhao05}
\bibinfo{author}{\bibfnamefont{L.}~\bibnamefont{Zhao}},
  \bibinfo{author}{\bibfnamefont{Y.-C.} \bibnamefont{Lai}},
  \bibinfo{author}{\bibfnamefont{K.}~\bibnamefont{Park}}, \bibnamefont{and}
  \bibinfo{author}{\bibfnamefont{N.}~\bibnamefont{Ye}},
  \bibinfo{journal}{Physical Review E} \textbf{\bibinfo{volume}{71}},
  \bibinfo{pages}{026125} (\bibinfo{year}{2005}).

\bibitem[{\citenamefont{Tadi\'{c} et~al.}(2006)\citenamefont{Tadi\'{c},
  Rodgers, and Thurner}}]{Tadic06}
\bibinfo{author}{\bibfnamefont{B.}~\bibnamefont{Tadi\'{c}}},
  \bibinfo{author}{\bibfnamefont{G.~J.} \bibnamefont{Rodgers}},
  \bibnamefont{and} \bibinfo{author}{\bibfnamefont{S.}~\bibnamefont{Thurner}}
  (\bibinfo{year}{2006}), \bibinfo{note}{physics/0606166}.

\bibitem[{\citenamefont{Erd\H{o}s and R\'{e}nyi}(1959)}]{Erdos59}
\bibinfo{author}{\bibfnamefont{P.}~\bibnamefont{Erd\H{o}s}} \bibnamefont{and}
  \bibinfo{author}{\bibfnamefont{A.}~\bibnamefont{R\'{e}nyi}},
  \bibinfo{journal}{Publicationes Mathematicae} \textbf{\bibinfo{volume}{6}},
  \bibinfo{pages}{290} (\bibinfo{year}{1959}).

\bibitem[{\citenamefont{Barab\'{a}si and Albert}(1997)}]{Barabasi97}
\bibinfo{author}{\bibfnamefont{A.-L.} \bibnamefont{Barab\'{a}si}}
  \bibnamefont{and} \bibinfo{author}{\bibfnamefont{R.}~\bibnamefont{Albert}},
  \bibinfo{journal}{Science} \textbf{\bibinfo{volume}{286}},
  \bibinfo{pages}{509} (\bibinfo{year}{1997}).

\end{thebibliography}

\end{document}